\documentclass[preprint,10pt]{aastex}

\newcommand{\fin}{\phi_{\rm in}}

\usepackage{graphicx}
\usepackage{amsmath,amssymb}
\usepackage{natbib}

\begin{document}

\title{Strong field effects on pulsar arrival times: circular orbits
  and equatorial beams}
\author{Yan Wang}\affil{Department of Astronomy,
Nanjing University, Nanjing 210093, China
and
Center for Gravitational Wave Astronomy
  and Department of Physics and Astronomy, University of Texas at
  Brownsville, Brownsville, Texas 78520
}

\author{Frederick A.~Jenet, Teviet Creighton, and Richard H. Price}
\affil{Center for Gravitational Wave Astronomy and
Department of Physics and Astronomy, University of Texas at Brownsville,
Brownsville, Texas 78520
}

\begin{abstract}
  If a pulsar orbits a supermassive black hole, the timing of pulses
  that pass close to the hole will show a variety of strong field
  effects.  To compute the intensity and timing of pulses that have
  passed close to a nonrotating black hole we introduce here a simple
  formalism based on two ``universal functions,'' one for the bending
  of photon trajectories and the other for the photon travel time on
  these trajectories. We apply this simple formalism to the case of a pulsar
  in circular orbit that beams its pulses into the orbital plane.  
In addition to the 
``primary'' pulses that reach the receiver by a more-or-less direct
  path, we
find that
  there are secondary and higher order pulses. These are
usually much dimmer than the  
primary pulses, but they can  be of comparable or even
  greater intensity if they are emitted when pulsar is on the
  side of the hole furthest from the receiver. We show  that there
  is a phase relationship of the primary and secondary pulses that is
  a probe of the strongly curved spacetime geometry. Analogs of these
  phenomena are expected in more general configurations, in which a
  pulsar in orbit around a hole emits pulses that are not confined to
  the orbital plane.
\end{abstract}
\maketitle

\section{Introduction}\label{sec:intro} 

Pulsars are rotating neutron stars that emit beams of radiation that
can be detected on Earth.  As the star rotates, this beam sweeps past
the Earth, producing regular pulses.  The timing of these pulses is
tied to the large inertial moment of a compact body, making it an
extremely stable clock: pulsars have been found whose pulse arrival
times fluctuate by less than 200
nano-seconds~\citep{2008ApJ...679..675V}.  Pulsar timing is thus an
excellent probe of delicate phenomena, including gravity, in the
vicinity of the pulsar.  It has been used to detect the tug of planets
orbiting a pulsar, to observe the gradual loss of orbital energy to
gravitational waves in binary systems containing a pulsar, and has
been proposed as a method of detecting cosmological gravitational
waves \citep{EW75,Sazhin78,Detweiler79}. In this paper we explore how
pulsar timing might probe the gravitational environment of pulsars
orbiting supermassive black holes.

Supermassive black holes are now thought to be ubiquitous in the
Universe, residing in the cores of most large galaxies, including our
own.  They have masses in the range of millions to billions of Solar
masses: our own Galaxy's black hole has an estimated mass of
$4\times10^6M_\odot$~\citep{2005ApJ...620..744G,2005ApJ...628..246E,2008arXiv0808.2870G}.
The best estimates of its mass come from observing the orbits of
O-type stars in the Galactic nucleus.  These observations have also
revealed that the Galactic nucleus is home to a significant population
of young massive stars, contrary to earlier
expectations \citep{2006JPhCS..54..279L}.  Perhaps the best model for
this population has it forming \textit{in situ} from the dense
molecular hydrogen disk surrounding the black hole, with an initial
mass function that is strongly tilted towards massive
stars \citep{2005MNRAS.364L..23N,2007MNRAS.374..515L,2007ApJ...669.1024M}.
The Galactic nucleus is thus a likely environment for neutron stars to
form, some of which may be pulsars orbiting deep within the potential
well of the supermassive black hole.  Although no such pulsars have
been detected, they may be discovered by future searches for periodic
signals with corrections for high accelerations (due to gravity)
and/or dispersion (due to plasma in the Galactic core).

The systems of interest here will consist of neutron stars at
distances down to within several million kilometers (a few
Schwarzschild radii) of the central black hole.  Even at these close
separations, it would take many years for the neutron star orbit to
decay due to gravitational radiation emission, making it possible to
conduct lengthy timing observations of a pulsar deep within the strong
field of a black hole.  (By contrast, a pulsar orbiting within a
thousand Schwarzschild radii of a $10M_\odot$ black hole would decay
within a year.)

In this paper we consider a pulsar to be in orbit around a
supermassive black hole, and we ask what effects the strong field of
the hole would have on pulsar observations.  Such effects would depend
on the details of the hole/pulsar system, and there are many details:
the pulsar orbital elements, the alignment of the pulsar spin axis and
the orbital plane, the pulsar spin rate, the angle between the pulsar
spin axis and the pulse emission direction, and the black hole
spin. (The black hole mass can be treated as a scaling parameter.)

An important step in understanding the details of strong field effects
on pulsar obsrvations is to understand them in simple cases. This is
what we do here, in two stages. First, we consider pulses emitted from
a pulsar in the neighborhood of a Schwarzschild (nonrotating) black
hole, with no restrictions on the pulsar orbit, spin, or emission
direction. We point out that in the case of a spherically symmetric
hole it is not necessary to do extensive computing of null
geodesics. Rather, it is necessary only to compute two functions of
the emission direction, one representing the bending of the light
path, and the second the time delay along that path. These two
``universal functions'' of emission direction are parameterized only by
the distance of the pulsar from the hole at the emission event.

The second stage in our analysis is to apply these universal functions
to a particularly simple astrophysical scenario, that of a pulsar
emitting in the orbital plane as it travels in a circular orbit around
a Schwarzschild hole. It turns out that even in this simplest possible
case, there is a rich set of interesting, and potentially observable
phenomena. Analogs of these phenomena could occur also in binary
pulsar systems and in binary neutron star/black hole systems of
comparable mass.

The paper is organized as follows. In Sec.~\ref{sec:unversal} we
define the two universal functions, and we present quantitative
results for them. Next, in Sec.~\ref{sec:TOAs}a we show how we can use
these functions to find any timing effect in the Schwarzschild
geometry. We then, in Sec.~\ref{sec:TOAs}b, apply this method
explicitly to pulses beamed in or near the orbital plane.  These
results are then applied, in Sec.~\ref{sec:examples}, to the simple
case of a circular orbit and equatorial beaming. In particular, we
show effects that would appear to a distant receiver of the pulse
trains.  A summary and a consideration of future applications of this
method are given in Sec.~\ref{sec:conc}. To allow the main ideas of
the paper to be as clear as possible, several sets of details have
been relegated to appendices.  Appendix A gives the details of finding
the direction of emission as a function of time for a pulsar in a
relativistic circular orbit.  Appendix B gives the details of the
computation of the universal functions.  The relativistic calculations
presented use the notational conventions of the text by Misner et
al. \citep{MTW}; in particular we use $c=G=1$.

\section{The universal functions for deflection and timing}\label{sec:unversal}

%%%%%%%%%%%%%%%%%%%%%%%%%%%%%%%%%%%%%%%%%%%%%%
\begin{figure}[h]
\begin{center}
\includegraphics[width=.6\textwidth]{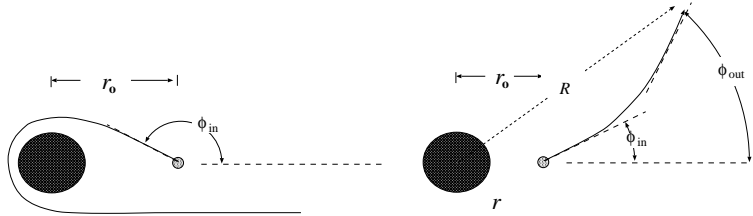}
\caption{Photon path from radial coordinate $r_0$ to $R$.
The path on the right indicates mild bending of the path
due to spacetime curvature. On the left is shown the highly
bent photon 
path which has an impact parameter near the critical value 
of photon capture, and a $\phi_{\rm out}$, at large $R$, of  $2\pi$.
 }
\label{fig:path}\end{center}
\end{figure}
%%%%%%%%%%%%%%%%%%%%%%%%%%%%%%%%%%%%%%%%%%%%%%

In a Schwarzschild hole with spacetime metric
\begin{equation}
  \label{eq:schw}
  ds^2=-\left(1-2M/r\right)dt^2+\left(1-2M/r\right)^{-1}dr^2+r^2\left(\
d\theta^2+\sin^2\theta\;d\phi^2
\right)\,,
\end{equation}
we consider a point at $r=r_0$, $\phi=0$, in the $\theta=\pi/2$
equatorial plane. At this point, a photon (i.e.\,, a null geodesic)
with 4-momentum $p^\mu$ is emitted in the equatorial plane.
From the fact that $p_\phi$ and $p_0$ are constants of motion
for the equatorial photon, it follows that the photon orbit
obeys
\begin{equation}
  \label{eq:photorbit}
  \frac{1}{r^4}\left(\frac{dr}{d\phi}\right)^2+\frac{1-2M/r}{r^2}=\frac{1}{b^2}\,,
\end{equation}
where $b$ is the photon impact parameter.

We define $\phi_{\rm in}$ to be $\tan^{-1}({r p^\phi/p^r})$, the angle
with respect to the outgoing direction, at which the photon is
emitted. (Note that this differs from the angle $\tan^{-1}({r
  p^\phi/\sqrt{g_{rr}\;}p^r})$ that would be measured in the local
frame of a coordinate stationary observer.) The value $\phi_{\rm in}$
is a constant characterizing the orbit, and is related to the impact
parameter $b$ by 
Eq.~(\ref{eq:photorbit}) with 
$r$  set equal to $r_0$:
\begin{equation}\label{eq:phineq}
\frac{1}{\tan^2{\phi_{\rm in}}}  +1-\frac{2M}{r_0}=\frac{r_0^2}{b^2}\,.
\end{equation}

A critical value of the impact parameter is $b_{\rm
  crit}=3\sqrt{3\;}M$; for $b<b_{\rm crit}$, inward going photons will be captured 
by the hole. For any $r_0$, capture will occur if $|\phi_{\rm in}|>\phi_{\rm crit}$
where $\phi_{\rm crit}$ is the root, between $\pi/2$ and $\pi$, of
\begin{equation}
  \label{eq:crit}
\tan{\phi_{\rm crit}} =-\,\left(\frac{r_0^2}{27M^2}-1+
\frac{2M}{r_0}\right)^{-1/2}
\,.
\end{equation}

For $|\phi_{\rm in}|$ less than $\phi_{\rm crit}$, our universal
functions relate the description of the photon at very large distance
$R$ to its emission conditions. The first of these functions measures
the bending of a photon trajectory, relative to the radially outgoing
direction from its emission point. To define this function we start by
defining ${\cal F}(\phi_{\rm in};R)$ to be the $\phi$ location, at
large radius $R$, of a photon emitted at radius $r_0$, at angle
$\phi_{\rm in}$ to the outgoing direction.  We then define
\begin{equation}
  \label{eq:Fdef}
  F(\phi_{\rm in})=\lim_{R\rightarrow\infty}{\cal F}(\phi_{\rm in};R )
\ .
\end{equation}
The function $F$ is an odd function of $\fin$; it is parametrized
by $r_0/M$, but has no other dependencies.
The second universal function $T$ is related to the coordinate time
${\cal T}(\phi_{\rm in};R)$
required for the the photon to reach asymptotically large $R$. To have
the result be a finite value, we subtract the coordinate time to reach
$R$ with $\phi_{\rm in}=0$, so that 
\begin{equation}
  \label{eq:Tdef}
  T(\phi_{\rm in})\equiv\lim\limits_{R\rightarrow\infty}\left[
{\cal T}(\phi_{\rm in};R)-{\cal T}(0;R)
\right]\,,
\end{equation}
and is an even function of $\fin$; like $F$, it is parametrized
by $r_0/M$, but has no other dependencies.

%%%%%%%%%%%%%%%%%%%%%%%%%%%%%%%%%%%%%%%%%%%%%%
\begin{figure}[h]
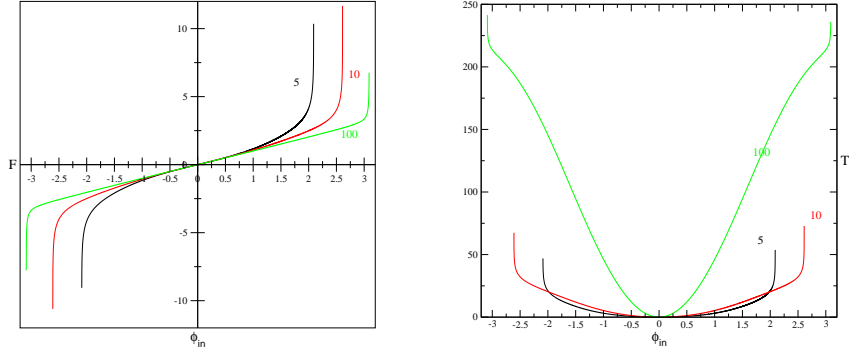
\begin{center}
\begin{center}
\includegraphics[height=.28\textwidth]{Ffun}\hspace{.4in}
\includegraphics[height=.28\textwidth]{Tfun}
\caption{Universal functions $F$ and $T$.  Curves are labeled with the
  values of $r_0/M$: 5, 10, and 100.  The values of $\phi_{\rm
    crit}$, from Eq.~(\ref{eq:crit}) are respectively 2.0895, 2.6109,
and 3.0896.
}
\label{fig:univfuns}\end{center}
\end{center}
\end{figure}
%%%%%%%%%%%%%%%%%%%%%%%%%%%%%%%%%%%%%%%%%%%%%%

Plots of $F$ and $T$ are given in Fig.~\ref{fig:univfuns} for three values 
of $r_0/M$. Approximate fits to these curves, good both for small $\phi_{\rm in}$
and for $\phi_{\rm in}$ near $\phi_{\rm crit}$, are given by
\begin{equation}\label{Fapprox}
F(\phi_{\rm in}) =
  -{\rm sign}(\phi_{\rm in})\log\left(1-|
    \frac{\phi_{\rm in}|}{\phi_{\rm crit}}\right)
  -\frac{\phi_{\rm in}}{\phi_{\rm crit}}+\phi_{\rm in}
\end{equation}

\begin{equation} \label{Tapprox}
  T(\phi_{\rm in})=r_{0}\left(1-\cos(\phi_{\rm in})\right)
  -{3\sqrt{3}}\log
\left(1-\frac{|\phi_{\rm in}|}{\phi_{\rm crit}}\right)
  -\frac{3\sqrt{3}}{\phi_{\rm crit}} |\phi_{\rm in}|\,. \\
%  -0.750006 \phi_{\rm in}^{2}+0.800266 \phi_{\rm in}^{4}-0.112303 \phi_{\rm in}^{6}
\end{equation}
The details of the computation of these curves, and more accurate analytic 
approxmations to the curves, are given in Appendix B.

\section{Pulse arrival times from universal functions }\label{sec:TOAs} 

\subsection{General case}

\begin{figure}[h]
\begin{center}\includegraphics[height=.27\textwidth]{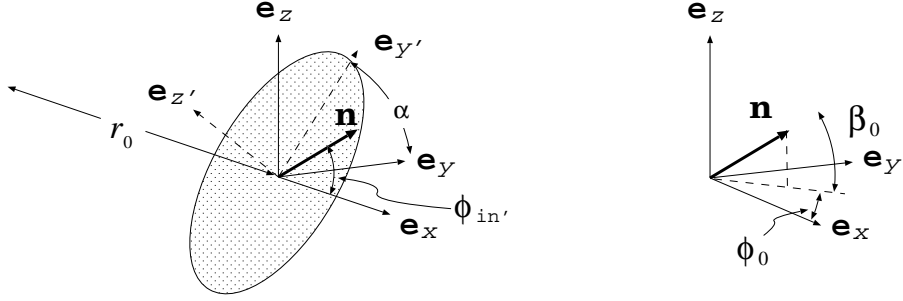}
\caption{Triads and angles used to define photon directions.
}
\label{fig:alphaplane}\end{center}
\end{figure}

We consider an emission event, at Schwarzschild radial coordinate
$r_0$, and a photon emitted in the direction ${\mathbf n}$. We let
$\alpha$ be the angle, as shown in Fig.~\ref{fig:alphaplane}, between
the orbital plane and the plane containing the photon trajectory,
i.e.\,, the plane defined by ${\mathbf n}$ and the radially outward
direction. (If these directions coincide, we take $\alpha$ to be zero.)

We introduce a spatial triad of orthonormal vectors ${\mathbf e}_{x},
{\mathbf e}_{y}, {\mathbf e}_{z}$, with ${\mathbf e}_{x}, {\mathbf
  e}_{y}$ in the pulsar orbital plane, and we define ${\mathbf
  e}_{x'}, {\mathbf e}_{y'}, {\mathbf e}_{z'}$, to be the orthonormal
spatial triad that result from a rotation by $\alpha$ around the
outgoing direction ${\mathbf e}_{x}$, so that ${\mathbf e}_{x'}, {\mathbf
  e}_{y'}$ lie in the plane of the photon trajectory.

\bigskip

To specify the direction of ${\mathbf n}$ we use the spherical polar
angles $\phi_0$, $\beta_0$ with respect to the ${\mathbf e}_{x},
{\mathbf e}_{y}, {\mathbf e}_{z}$ directions. (Note that for
convenience we are using the latitude $\beta$, rather than the more
typical colatitude $\theta=\pi/2-\beta$.)
From the rotational transformation between the unprimed and primed triads
we get
\begin{equation}\label{alphaeq}
  \tan\alpha=\frac{\tan\beta_0}{\sin\phi_0}
\end{equation}
\begin{equation}
  \cos\phi_{\rm in'}=\cos\beta_0\;\cos\phi_0\;.
\end{equation}

%%%%%%%%%%%%%%%%%%%%%%%%%%%%%%%%%%%%%%%%%%%%%%
%%%%%%%%%%%%%%%%%%%%%%%%%%%%%%%%%%%%%%%%%%%%%%

Here $\phi_{\rm in'}$ is the angle, in the plane of the photon
trajectory, between the photon direction and the radially outgoing 
direction. We can, therefore, find $\phi_{\infty'}$, the asymptotic $\phi$
direction, from 
\begin{equation}\label{Feq}
  \phi_{\infty'}=F(\phi_{\rm in'})\ .
\end{equation}
The spherical polar angles, in the 
${\mathbf e}_{x},
{\mathbf e}_{y}, {\mathbf e}_{z}$ frame,
for the final photon direction, are then
\begin{equation}
  \cos\theta_\infty=\sin\alpha\sin\phi_{\infty'}\quad\quad\quad\quad
\tan{\phi_{\infty}}=\cos\alpha\tan{\phi_{\infty'}}\,.
\end{equation}
We summarize here the set of equations that give the final photon
direction $\beta_{\infty},\phi_{\infty}$, and relative time of arrival
$t_{\infty}$, in terms of the original photon direction
$\beta_{0},\phi_{0}$:
\begin{eqnarray}
  \cos\phi_{\rm in'}&=&\cos\beta_0\;\cos\phi_0  \label{eq:summary1}
\\
    \sin\beta_\infty&=&\sin{\big(F(\phi_{\rm in'})\big)}\,\frac{\sin\beta_0}
  {
    \sqrt{\sin^2\beta_0+\cos^2\beta_0\sin^2\phi_0 
      \;}}\\
    \tan\phi_\infty&=&\tan{\big(F(\phi_{\rm in'})\big)}\,
\frac{\cos\beta_0 \sin\phi_0}
  {
    \sqrt{\sin^2\beta_0+\cos^2\beta_0\sin^2\phi_0 
      \;}}\\
t_{\infty}&=&t_e+T(\phi_{\rm in'})\label{eq:summary4}\ .
\end{eqnarray}

\subsection{Special case: beaming in orbital plane}

If the pulsar beam is emitted into the orbital plane, then $\beta_0=0$
and the relations in Eqs.~(\ref{eq:summary1})-(\ref{eq:summary4})
reduce to $\phi_{\rm in'}=\phi_0$, and $\phi_{\infty}=F(\phi_{\rm in})$.
The photon remains in the orbital plane, so $\beta_\infty=\beta_0=0$.

For the simplest example of strong field effects we restrict ourselves 
to the case in which the receiving antenna is precisely in the orbital plane
of the pulsar. This means that photons reaching the antenna must have 
exceedingly small values of $\beta_0$ (of order of the antenna size divided
by tens of kiloparsecs).
In this case the initial-to-final
transformations in Eqs.~(\ref{eq:summary1})-(\ref{eq:summary4}) can be 
replaced by approximations valid to first order in $\beta_0$:
\begin{eqnarray}
   \beta_\infty&=&\beta_0\; 
\frac{\sin\left(F(\phi_0)\right)}{\sin\phi_0}\label{smallbeta1}\\
\phi_\infty&=&F(\phi_0)\label{smallbeta2}\,.
\end{eqnarray}
In these equations it is assumed that $\beta_0\ll\phi_0$, a condition
that will be violated only by photons directed very nearly radially 
outward, and hence experiencing no strong field effects.
The sine function in the numerator of Eq.~(\ref{smallbeta1})  
implies $\beta_\infty=0$ if $F(\phi_0)=\pi$. This  
corresponds to the case that all small-$\beta_0$ photons with the same value
of $\phi_0$ are focused into the orbital plane, creating an intensity
amplification that would be infinite aside from diffraction limitation.

Equations (\ref{smallbeta1}), (\ref{smallbeta2}) give us the factors
by which strong field effects lead to 
convergence or divergence of the photon beam both in the pulsar orbital 
plane
$f_{\rm plane}$, and perpendicular to it $f_{\rm perp}$:
\begin{equation}
  \label{eq:fs}
  f_{\rm plane}=\frac{dF}{d\phi_0}\quad\quad\quad f_{\rm perp}=
\frac{\sin\left(F(\phi_0)\right)}{\sin\phi_0}\,.
\end{equation}
The total strong-field effect on the cross sectional area of the beam
reaching the receiver is $f_{\rm plane}f_{\rm perp}$. The effect on
the photon flux at the receiver will therefore be $1/ \left(f_{\rm
    plane}f_{\rm perp}\right)$.  The total radio intensity reaching
the receiver will also depend on the gravitational redshift of the
photons and the Doppler shift due to the motion of the emitting
pulsar. Both effects are of order $M/r_0$.

\section{Appearance and timing of pulses
}\label{sec:examples}

We make another simplifying assumption: that the pulsar that emits its
beam in the equatorial plane, is moving in a circular orbit at radius
$r_0$. For definiteness we take the angular location of the pulsar to
be given by $\phi_{\rm orb}=\Omega t$, where ``$t$'' here and below, indicates
coordinate time.  At a particular emission time
$t_{\rm e}$, it is shown in Appendix \ref{sec:appA} that the angle at which the
photon is emitted is given by
\begin{equation}
  \label{eq:tanphiin}
\tan{\big(\phi_{\rm in}(t_{\rm e})\big)}=
\frac{r_0\Omega\left(1-2M/r_0\right)^{-1/2}
+
\cos{({\gamma^{-1}[\Omega-\omega]}t_{\rm e}+\delta)}}
{\gamma^{-1}\sin({\gamma^{-1}[\Omega-\omega]} t_{\rm e}+\delta)
}
\,.
\end{equation}
Here $r_0$ is the  radial coordinate of the circular 
orbit; $\Omega=\sqrt{M/r_0^3\;}$  is the pulsar orbital angular velocity
(per unit coordinate time); $\omega$ is the pulsar spin rate as measured
by a comoving observer; 
the Lorentz factor $\gamma$ is $1/\sqrt{1-3M/r_0\;}$; and
$\delta$ is a phase constant specifying the 
direction of the beam at $t_{\rm e}=0$.

If a distant radio receiver is at $\phi_{\rm rec}$, then to find the emission time
of photons that are destined to be received, we must solve
\begin{equation}
  \label{eq:eqforte}
  \Omega t_{\rm e}+F\big(
\phi_{\rm in}(t_{\rm e})\big)=\phi_{\rm rec}\,.
\end{equation}
Due to the nature of $F$ near the critical angles $\phi_{\rm crit}$,
there are, in principle an infinite number of solutions corresponding
to photon orbits that circle the gravitating center zero times, once,
twice, etc. It will be useful to refer to these received pulses as
primary, secondary (once around the gravitating center), tertiary, and so
forth. We shall see, however, that these distinctions can become ambiguous.

The observationally important questions are the timing and appearance
of the pulses that show strong field effects.  We start by arguing
that the effect on the pulse shape will be negligible if the rotation
period (seconds or less) of the pulsar is much smaller than the pulsar
orbital period (1000s of seconds to years, depending on $r_0/M$ and on
the mass of the central supermassive black hole).
%%%%%%%%%%%%%%%%%%%%%%%%%%%%%%%%%%%%%%%%%%%%%%
\begin{figure}[h]
\begin{center}\includegraphics[height=.2\textwidth]{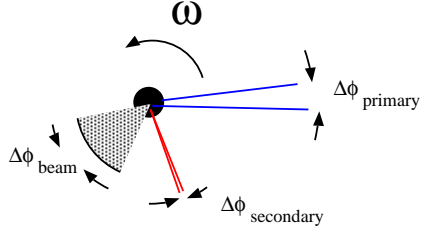}
\caption{
The pulsar beam sweeping through the narrow range of directions
that connect to the receiver.
}
\label{fig:detail}\end{center}
\end{figure}
%%%%%%%%%%%%%%%%%%%%%%%%%%%%%%%%%%%%%%%%%%%%%%

Figure \ref{fig:detail} shows the geometry of beam emission, with
$\Delta\phi_{\rm beam}$ indicating the inherent angular width (in the
orbital plane) of the pulsar beam, and with $\Delta\phi_{\rm primary}$
and $\Delta\phi_{\rm secondary}$ indicating (but exaggerating) the
receiver acceptance directions, i.e., the
range of photon directions that connect to the distant radio
receiver. The angular size of $\Delta\phi_{\rm primary}$ is
approximately the ratio of the receiver diameter to the receiver
distance. The much smaller angular size $\Delta\phi_{\rm secondary}$
is further divided by $dF/d\phi_{\rm in}$, a large number. For
tertiary and subsequent beams, the same description applies except
that the value of $dF/d\phi_{\rm in}$ is even larger. 

If the pulsar were at a fixed coordinate position, and rotating at
$\omega$, then the shape of the received pulse can be viewed as the
result of the narrow cones of receiver acceptance sweeping through the
$\Delta\phi_{\rm beam}$ beam profile. This viewpoint makes it clear
that the time profile of the primary, secondary, etc. beams would have
the same shape.  The pulsar, of course, is not coordinate stationary,
but is orbiting with orbital speed $\Omega$. This means that there
will be a small change in the nature of $dF/d\phi_{\rm in}$ during the
passage of the beam width through the receiver acceptance cones, since
the value of $\phi_{\rm in}$ for received photons will change slightly
during beam reception. To estimate this effect we can consider the
change $\Delta F$ during the change in pulsar orbital location
$\Delta\phi_{\rm orb}$:
  \begin{equation}
    \frac{dF}{d\phi_{\rm in}}\,\Delta\phi_{\rm in}\approx
    \frac{dF}{d\phi_{\rm in}}
\,\Delta\phi_{\rm orb}
\frac{\Omega}{\omega}\,.
\approx 
e^{F}\,2\pi\,\frac{\Omega}{\omega}\,.
\end{equation}
(The second approximations assumes that $\phi_{\rm in}\approx\phi_{\rm
  crit}$ and that $F$ is large.)  For a secondary beam, $F$ must be of
order $2\pi$; for a tertiary beam, $F$ must be order $4\pi$ etc. Thus,
in the case of high order beams or for exceptionally large 
values of $\Omega/\omega$, there could be some distortion of received
pulse shapes. This possibility will not be considered further in the
current paper.

%%%%%%%%%%%%%%%%%%%%%%%%%%%%%%%%%%%%%%%%%%%%%%
\begin{figure}[h]%FIG 5
\begin{center}
\includegraphics[height=.4\textwidth]{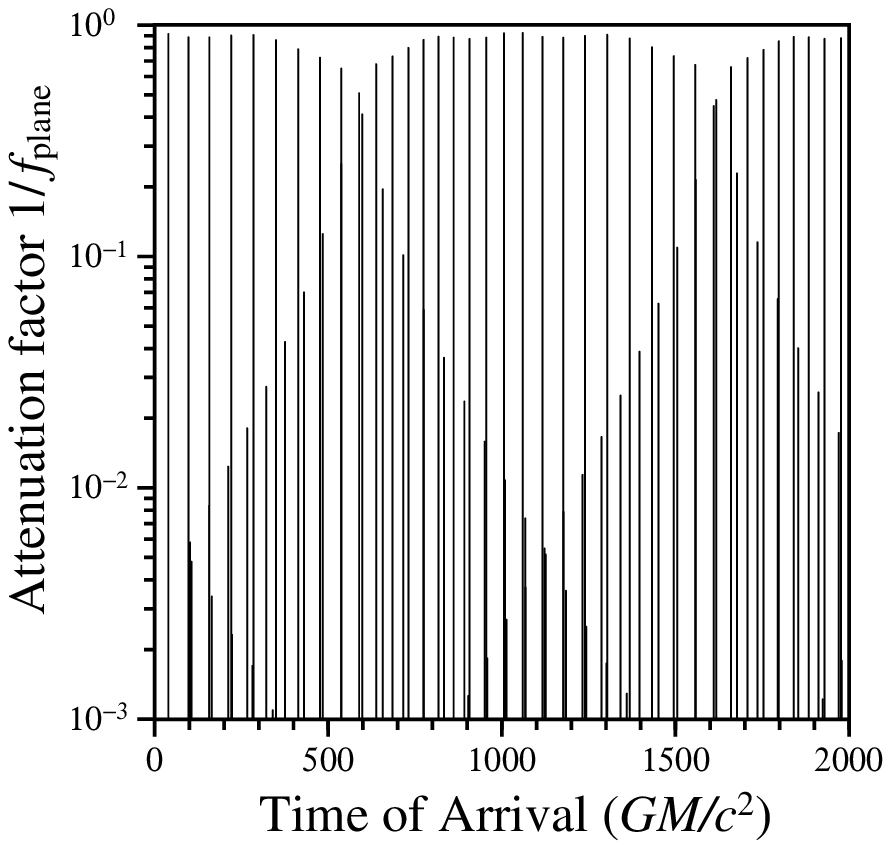}
\includegraphics[height=.4\textwidth]{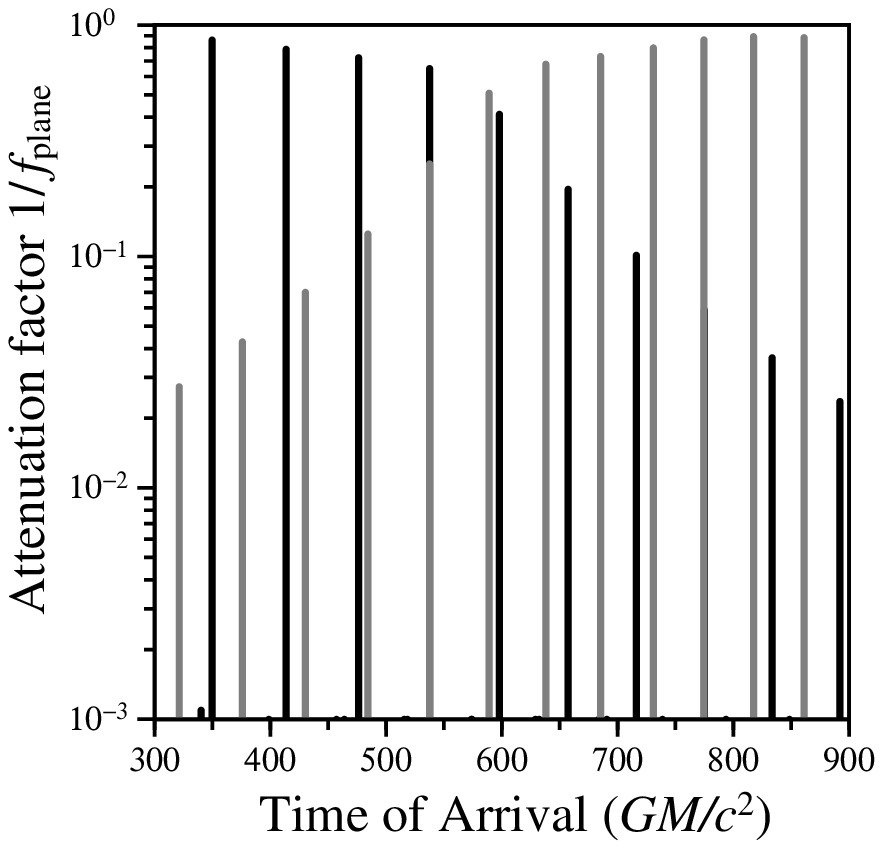}
\caption{
The factor $1/f_{\rm plane}$ for photon capture events from 
a pulsar with orbital radius $30\times GM/c^2$, and pulsar 
rotation rate  20.2 times the orbital rotation rate.
Events for several orbits are shown on the left. Details
are shown on the right for the orbital epoch during which
the pulsar is on the far side of the hole.
}
\label{fig:amplitude30}\end{center}
\end{figure}
%%%%%%%%%%%%%%%%%%%%%%%%%%%%%%%%%%%%%%%%%%%%%%

Figure~\ref{fig:amplitude30} shows a schematic of pulses emitted from
a pulsar orbiting at a distance $30\times GM/c^2$ from a supermassive
black hole of mass $M$.  The pulsar rotation rate has been chosen to
be very low, only 20.2 times the orbital frequency, in order to show
clearly some of the phenomenology of the pulse arrival times.  For a
$4\times10^6 M_\odot$ black hole, this means an orbital period of
$2\pi\times30^{3/2}\times GM/c^3\approx20,000$~seconds and a pulsar
rotation period around 1000~seconds; an actual observed pulsar would
likely be rotating hundreds or thousands of times faster.  The
horizontal axis represents the pulse arrival time (relative to an
arbitrary start time), while the vertical axis gives the attenuation
factor $1/f_\mathrm{plane}$ due to horizontal spreading of the beam.
Using this factor as a tag on the pulses is useful in the discussion
of the pulse sequences.  The factor gives a rough indication of
relative pulse strengths, though the full amplification or attenuation
of the beam requires the complete spreading factor $1/
f_\mathrm{plane} f_\mathrm{perp}$, and will be discussed below.

%%%%%%%%%%%%%%%%%%%%%%%%%%%%%%%%%%%%%%%%%%%%%%
\begin{figure}[h]%FIG 6
\begin{center}\includegraphics[height=.17\textwidth]{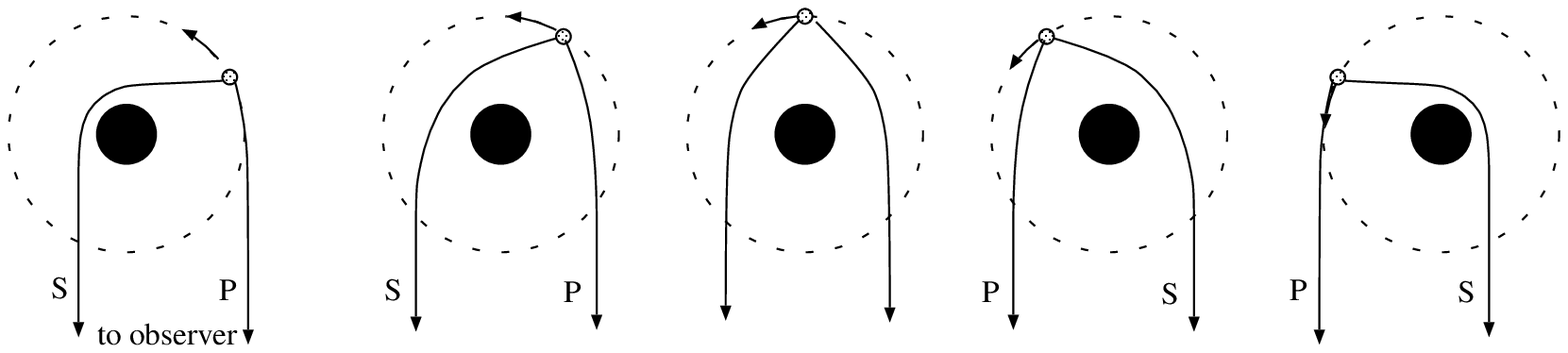}
\caption{Prograde and retrograde photon trajectories.
The pulsar is shown orbiting in the counterclockwise direction around
a much more massive black hole.
The leftmost cartoon shows that the observing radio telescope
receives a primary (P) direct pulse and a highly bent secondary (S) pulse.
The subsequent panels show trajectories as the pulsar continues its
orbtial motion. The primary trajectory has increased bending and the 
secondary less, until the pulsar is directly opposite the receiver and 
the two trajectories are symmetric. The following panels shows how 
the prograde trajectory then becomes the primary (less bent) trajectory.
}
\label{fig:proretro}\end{center}
\end{figure}
%%%%%%%%%%%%%%%%%%%%%%%%%%%%%%%%%%%%%%%%%%%%%%

Initially, the taller set of pulses represents pulses that arrive at
the detector from the pulsar along a more-or-less direct path.  The
initially weaker but strengthening set represents pulses that arrive
from a path that is bent around the black hole in a sense prograde to
the orbit. Figure~\ref{fig:proretro} illustrates how this secondary
path ``unwinds'' and becomes more direct as the pulsar moves along its
orbit.  The effective photon path length $\ell$ also shortens with
time, giving rise to a shorter (blueshifted) pulse period
$P=P_0(1+\dot{\ell}/c)$.  Around time $600\times GM/c^3$, the pulsar
passes behind the black hole (the middle panel in
Fig.~\ref{fig:proretro}).  After this, the prograde path becomes
the more direct path, while what was formerly the more direct path now
gets wound around the black hole in the retrograde sense, causing its
pulses to weaken and the period of the pulses to redshift.  The pulse
timing near the time of  this orbital phase is
shown in greater detail in the second panel of Figure~\ref{fig:amplitude30},
where the prograde-wound pulses are shown in gray to distinguish them
from the other pulses.  The process repeats itself one orbit later.
The ragged pulses at the bottom of the two graphs are pulses from
paths with higher-order windings around the black hole; some of these
will eventually unwind to become, for a time, the most direct path.

%%%%%%%%%%%%%%%%%%%%%%%%%%%%%%%%%%%%%%%%%%%%%%
\begin{figure}[h]%FIG 7
\begin{center}
\includegraphics[height=.35\textwidth]{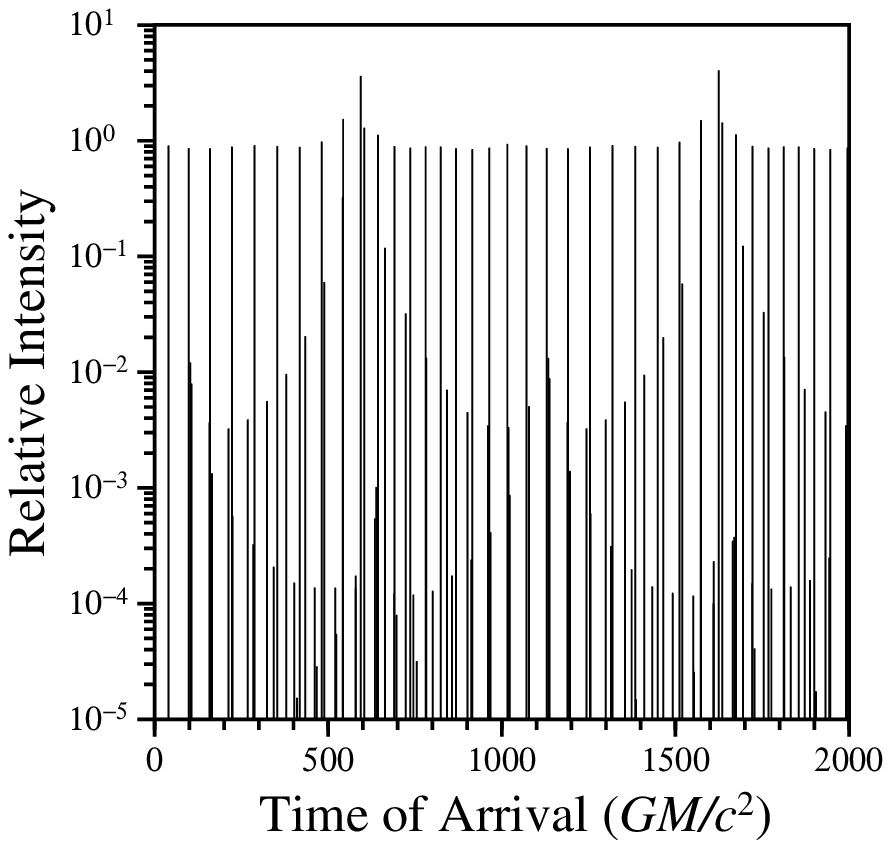}
\includegraphics[height=.35\textwidth]{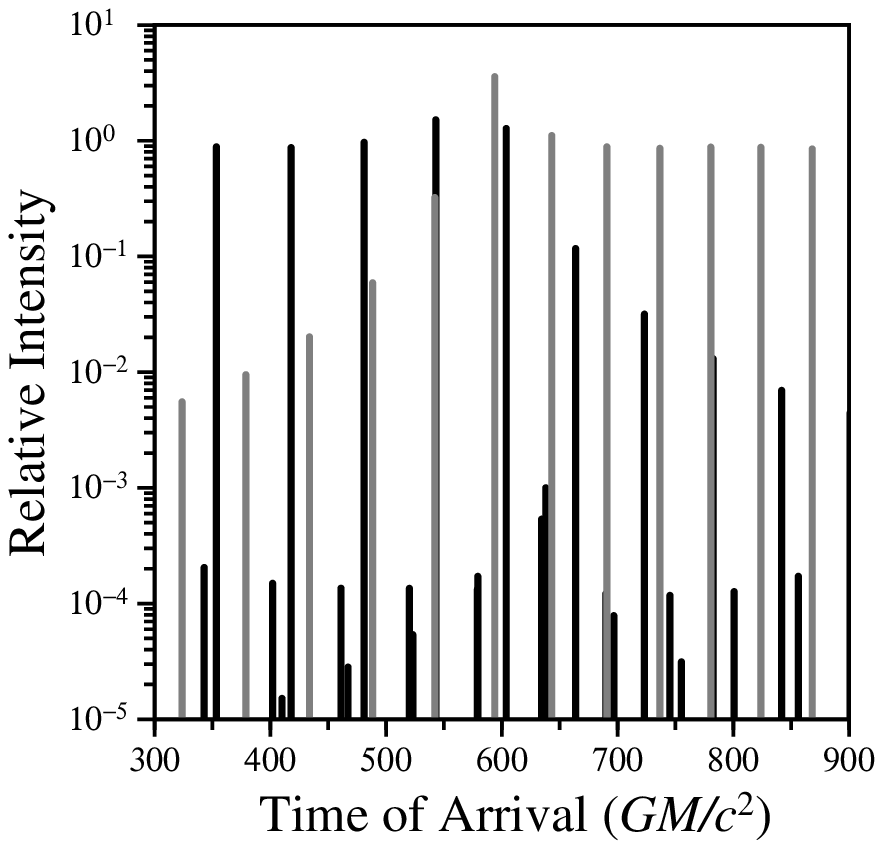}
\caption{The relative intensity of pulses for the same system shown 
in Fig.~\ref{fig:amplitude30}}
\label{fig:intensity30}\end{center}
\end{figure}
%%%%%%%%%%%%%%%%%%%%%%%%%%%%%%%%%%%%%%%%%%%%%%

Figure~\ref{fig:intensity30} shows the same system, but with the vertical
axis now showing the full pulse amplification (or attenuation) factor
$1/f_\mathrm{plane} f_\mathrm{perp}$.  
Around the same time as the
``primary'' and ``secondary'' pulse trains swap roles, both sets of
pulses go through a spike of amplification due to strong lensing
around the black hole.

It should be understood that the ``intensity'' in Figure~\ref{fig:intensity30}
and below refers to the photons received per unit time. The true energy 
intensity would include the effect of Doppler shifts due to the 
orbital motion. These effects, of order $r_0\Omega/c$, are
significant -- around 20\% for an orbit with $r_0=30\times GM/c^2$ -- but
are omitted to emphasize the photon path effects.

\begin{figure}[h]%FIG 8
\begin{center}
\includegraphics[height=.4\textwidth]{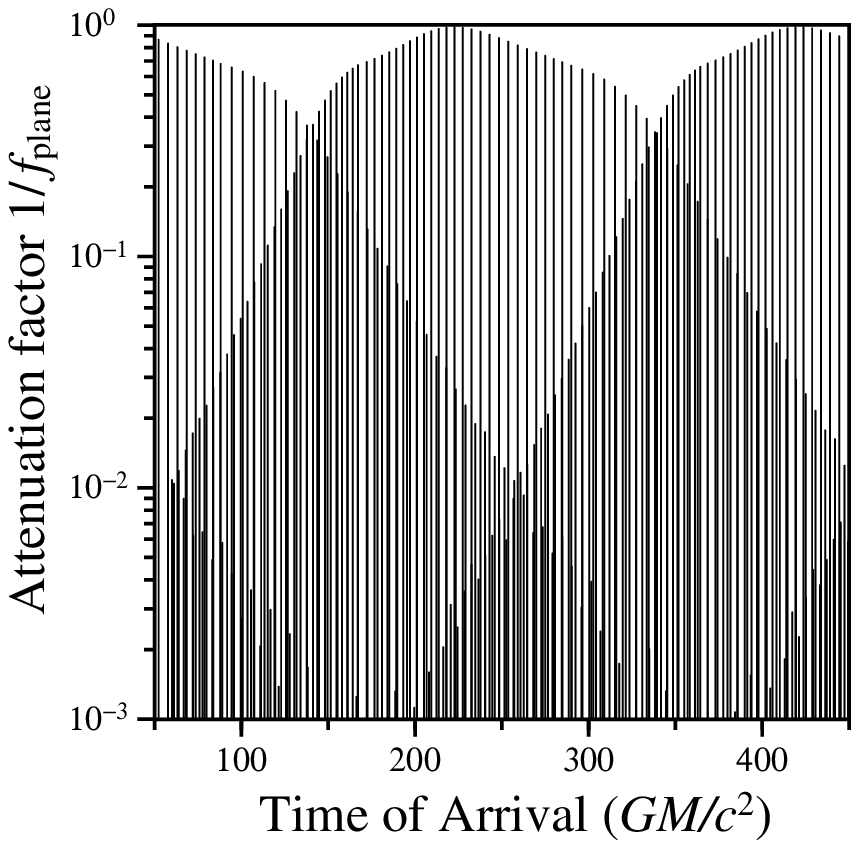}
\includegraphics[height=.4\textwidth]{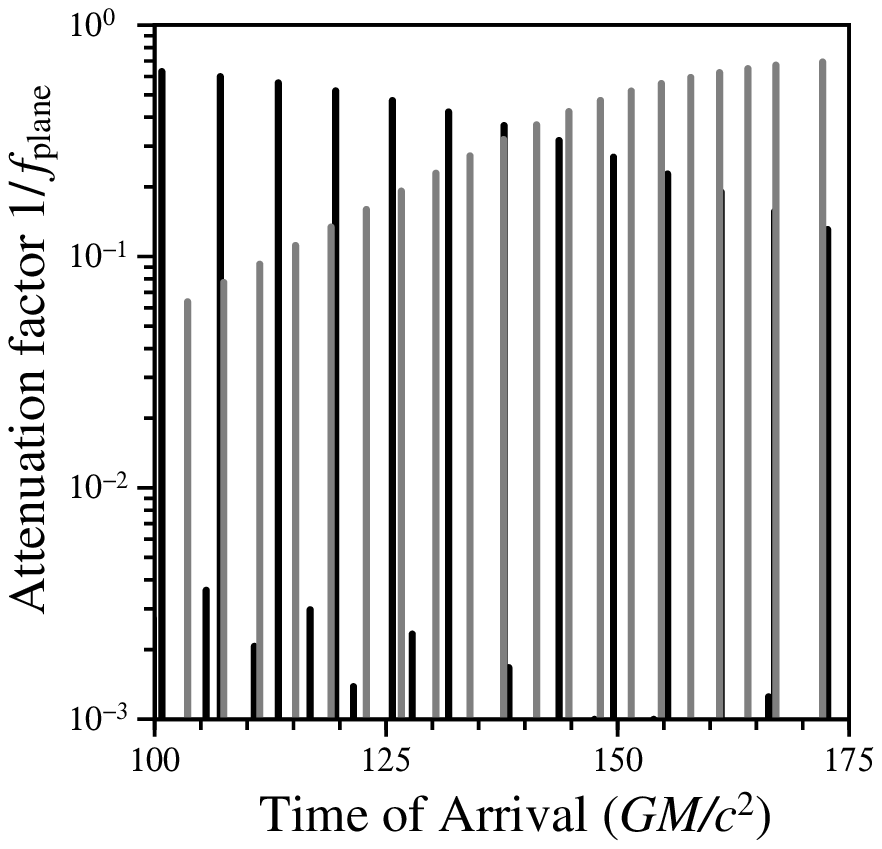}
\caption{The same plot as in Fig.~\ref{fig:amplitude30}, but for 
a more relativistic system, with 
with orbital radius $10\times GM/c^2$, and 
with rotation rate 47 times the orbital frequency.
}
\label{fig:amplitude10}\end{center}
\end{figure}

The details of the pulse timing and amplification phenomenology depend
on the orbital radius. To illustrate this, Fig.~\ref{fig:amplitude10}
shows the same information as Fig.~\ref{fig:amplitude30} but for a
more relativistic system, where the pulsar orbital radius is only
$10\times GM/c^2$.  Again, the pulsar rotation rate is taken to be
artificially slow (only 47 times the orbital frequency) in order for
the figure to show distinct pulses.  In this more relativistic system,
the pulse period shift, beaming asymmetry, and ``tertiary''
(multiply-wound) pulse paths are more apparent.

%%%%%%%%%%%%%%%%%%%%%%%%%%%%%%%%%%%%%%%%%%%%%%
%%%%%%%%%%%%%%%%%%%%%%%%%%%%%%%%%%%%%%%%%%%%%%

%%%%%%%%%%%%%%%%%%%%%%%%%%%%%%%%%%%%%%%%%%%%%%
\begin{figure}[h]%FIG 9
\begin{center}\includegraphics[height=.57\textwidth]{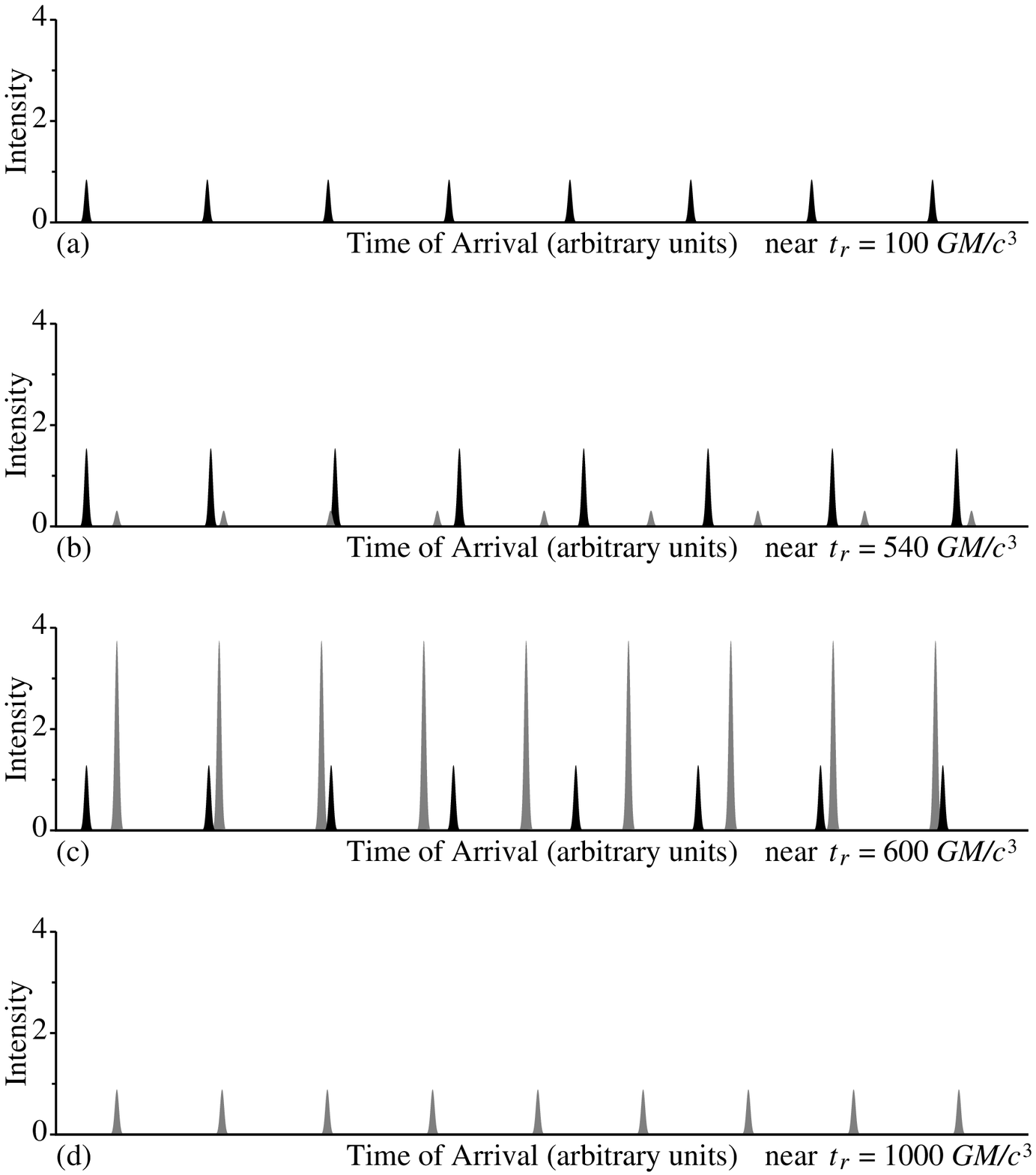}
\caption{The appearance of received radio pulses for several 
orbital epochs of a pulsar orbiting with radius 
$30\times GM/c^2$. The pulsar rotation rate is assumed 
to be much greater than the orbital frequency.
}
\label{fig:pulsetrain}\end{center}
\end{figure}
%%%%%%%%%%%%%%%%%%%%%%%%%%%%%%%%%%%%%%%%%%%%%%

Figure~\ref{fig:pulsetrain} shows the pulse trains that a radio
astronomer might actually see from the  $r_0=
30\times GM/c^2$
system of Figs.~\ref{fig:amplitude30} and \ref{fig:intensity30},
but assuming a more
realistic pulsar rotation rate that is thousands of times faster than
the orbital frequency.  For such a sufficiently fast rotation rate,
the pulsar orbital position changes negligibly during the emission of
a set of a few pulses.  The pulse intensity, determined by
$F(\phi_{\rm in})$, and $dF/d\phi_{\rm in}$, and the red/blue shift in
the period, determined by $\dot{\ell}$, therefore are fixed, once the
orbital phase is fixed. This means that we can show pulse intensity
and primary vs.~secondary phase shifts without specifying any one
pulsar rotation rate. For this reason, Fig.~\ref{fig:pulsetrain} does
not specify units of the horizontal axis; it is understood that the
spacing between direct pulses is the period of pulsar rotation. The
time that {\em is} specified for each panel is the time that
determines the orbital position when the set of pulses is emitted.

The vertical axis represents the pulse power relative to an
undeflected beam, and each pulse is given a Gaussian shape.  Panel~(a)
is a segment of the pulse train when the pulsar is on the near side of
the black hole, and only a single train of direct pulses is visible.
Panel~(b) is a segment from the orbital phase during which the pulsar
is passing behind the black hole: a secondary train of prograde-bent
pulses, with blueshifted period, starts to appear.  In panel~(c), the
prograde path has become the more direct path, and the pulses from
what was originally the direct path are weakening, and their period
redshifting, as the path becomes a bent retrograde one around the
black hole (though both sets of pulses are amplified due to lensing).
In panel~(d), the original pulse train has disappeared almost
entirely, leaving only the pulses from the newly-unwound path.

We note that when the most-direct path shifts from passing retrograde
around the hole to prograde around the hole (see
Fig.~\ref{fig:proretro}), there is a sudden jump in the frequency of
the strongest pulse train, as the derivative of the effective path
length $\dot{\ell}$ switches from lengthening to shortening.
(Equivalently, in this geometry, a train of high-frequency
``secondary'' pulses rises up and replaces the low-frequency
``primary'' pulses.)  This is the strong-field analog of the
well-known cusp in the Shapiro time delay curve, where the time delay
switches from increasing to decreasing; since the observed pulse
period is multiplied by 1 plus the time derivative of the time delay
$P_\mathrm{obs}=P_0(1+dT_\mathrm{Shapiro}/dt)$, this corresponds to a
discontinuous jump in period.  Unlike the case of the ``standard'' Shapiro
delay, the discontinuity is finite even for a perfect alignment, since
the deflecting mass is a black hole of finite size.  There is also a
slight asymmetry due to the special relativistic beaming (``headlight
effect'') of the pulsar in its orbit.

\section{Summary and Conclusions}\label{sec:conc}

We have introduced the problem of black hole effects on pulses from an
orbiting pulsar. We have shown that in the case of a spherically
symmetric (i.e., nonrotating) hole, the use of two ``universal
functions'' removes the need for extensive computation of null
geodesics. The universal function approach to computation has been
used for a first exploration of the phenomena that might be
encountered with pulses from a pulsar-hole system.  This first
exploration investigates pulse emission in the orbital plane, for a
pulsar in circular orbit around a nonrotating hole.

Even for this highly simplified configuration we have found a number
of interesting phenomena: (i)~``Primary'' pulses (pulses that travel
from the pulsar to the receiver with relatively little gravitational
bending) are continually accompanied by higher order (secondary,
tertiary,...) pulses emitted during earlier pulsar orbits. (ii)~As
would be expected, the primary pulses do not have a constant period,
but rather have a period that is modulated by orbital
motion. (iii)~The period of primary and of higher order pulses is not
precisely the same, thus the higher order pulses arrive with a phase
shift, relative to the primary pulses, that varies from one pulse 
to the next. (iv)~Strong field effects dominate the pulsar
observations when the pulsar is on the far side of the hole, the
side opposite that of the receiver. In this case the emitted
pulse can reach the receiver only by a highly bent path. (v)~During
the epoch of emission from the far side, the sequence of primary pulses
and the sequence of secondary pulses, exchange roles. One consequence
is that there is no clear distinction of primary and secondary 
for pulses emitted when the pulsar is close to the middle of its passage 
through the far side of the  hole. (vi)~For emission
during most of the pulsar orbit, the secondary pulses have much lower
intensity than the primary pulses. As the pulsar moves toward the dark
side, however, the secondary pulses increase in intensity and the
primary pulses decrease.  (vii)~Among the pulses emitted from the dark
side are pulses that are {\em amplified} by strong field effects
analogous to gravitational lensing. 

The simple configuration studied would be expected to exaggerate
strong field effects when compared to a more realistic configuration
in which pulse emission is directed well out of the orbital plane.
For example, if the pulsar spin axis is perpendicular to the orbital
plane (as in our simple configuration), and emission is not close to
the orbital plane (contrary to our simple configuration), the pulse
trajectory will never cross the orbital plane, and hence never come
sufficiently near the hole for strong field effects to be significant.
More interesting is the case in which the spin axis and beaming
details are such that the beam does cross (or pass close to) the
orbital plane.  Particularly noteworthy would be a configuration in
which the receiver receives no primary pulses, and receives pulses
only with the aid of strong field bending. It is also of note that
some of the phenomena we have described, especially the existence of
higher order pulses, can also occur in principle in binary pulsar
systems, and pulsar-hole binaries of comparable mass.

Work is underway on investigating configuratins with out of orbit
beaming.  An exploration of the large parameter space will be feasible
with the efficiency provided by the universal function approach, so we
are, at least at first, restriciting attention to nonrotating
supermassive holes.

\section{Acknowledgment} We gratefully acknowledge support by the
National Science Foundation under grants AST0545837, PHY0554367 and
0734800. We also thank the NASA Center for Gravitational Wave
Astronomy at University of Texas at Brownsville. YW acknowledges
support by the Chinese National Science Foundation under grant
10773005.

\appendix

\section{Emission direction as a function of time}\label{sec:appA}

%%%%%%%%%%%%%%%%%%%%%%%%%%%%%%%%%%%%%%%%%%%%%%
\begin{figure}[h]
\begin{center}
\includegraphics[width=.4\textwidth]{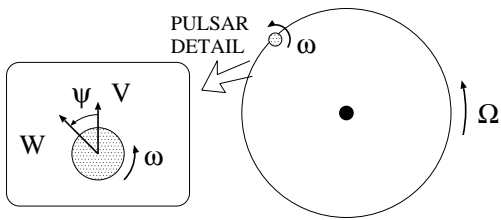}
\caption{The figure illustrates the orbital motion of the 
pulsar at frequency $\Omega$, the parallel transported 
spatial direction $V^\mu$, and the pulsar beam direction $W^\mu$
rotating at frequency $\omega$ relative to $V^\mu$, as observed 
in a frame comoving with the pulsar.
}
\label{fig:teviet}
\end{center}
\end{figure}
%%%%%%%%%%%%%%%%%%%%%%%%%%%%%%%%%%%%%%%%%%%%%%

In the orbital plane we let the angular position of the pulsar be
$\phi=\Omega t$, with $t$ the Scharzschild coordinate time and the
pulsar's 4-velocity components
are\begin{equation}
  \label{eq:Ucomps}
  U^0=\gamma\quad\quad U^\phi=\gamma\Omega
\end{equation}
where 
\begin{equation}
  \label{eq:gamma}
  \gamma=\frac{1}{
\sqrt{1-2M/r_0-r_0^2\Omega^2\;}}
=\frac{1}{
\sqrt{1-3M/r_0
\;}}\ .
\end{equation}

We let $V^\mu$ be a 4-vector that is spatial (that is, orthogonal to
$U^\mu$), that has no component out of the equatorial orbital plane,
and that is parallel transported around with the pulsar. (Since the
pulsar world line is a geodesic, this is the same as Fermi-Walker
transporting $V^\mu$.) It is straightforward to show that the
components of such an ``inertial direction'' are
\begin{equation}
  \label{eq:Vcomps}
V^r=\gamma^{-1}{r_0}K\sin{(\gamma^{-1}{\Omega}t+\delta)}
\quad\quad
V^\phi=K\cos{(\gamma^{-1}{\Omega}t+\delta)}
\quad\quad
V^0=\frac{r_0^2\Omega}{{1-2M/r_0\;}}K\cos{(\gamma^{-1}{\Omega}t+\delta)}\,.
\end{equation}
where $K$ is a scaling constant and $\delta$ is a phase factor
determining the direction in which $V^\mu$ is pointing at $t=0$. We
let $W^\mu$ be a spatial vector, in the equatorial plane, that points
in the direction of the pulsar beam.  We let $\psi$ be the angle,
measured in the positive sense (the positive sense for $\Omega$) from
$V^\mu$ to $W^\mu$. In terms of the proper time $\tau$ measured by the 
pulsar, we define the locally observed pulsar spin rate $\omega$ by 
\begin{equation}
  \label{eq:psioft}
  \psi=\omega\tau=\omega t/\gamma\,.
\end{equation}
From $W^\mu U_\mu=0$ and $\cos{\psi}=W^\mu V_\mu/VW$ we get
$$
W^r=\frac{A}{\gamma}\sin{(\gamma^{-1}{[\Omega-\omega]}t+\delta)}
\quad\quad
W^\phi=\frac{A}{r_0}\cos{(\gamma^{-1}{[\Omega-\omega]}t+\delta)}
$$
\begin{equation}
  \label{eq:Wcomps}
W^0=\frac{r_0\Omega A}{{1-2M/r_0\;}}
\cos{(\gamma^{-1}{[\Omega-\omega]}t+\delta)}\,.
\end{equation}

In the comoving frame of the pulsar the spatial direction of the photon
beam is $W^\mu$, thus the photon 4-momentum must have the form
$\vec{p}=\kappa\vec{U}+\vec{W}$.
The value of $\kappa$ follows from $p^\mu p_\mu=0$ and we find
\begin{equation}
  \label{eq:pform}
  \vec{p}=\frac{A}{\gamma\sqrt{1-2M/r_0\;}}\,\vec{U}+\vec{W}\,,
\end{equation}
from which we get the components
\begin{eqnarray}
  \label{eq:fourmomcomps}
p^0&=&\frac{A}{\sqrt{1-2M/r_0\;}} 
+\frac{r_0\Omega A}{{1-2M/r_0\;}}
\cos{(\gamma^{-1}{[\Omega-\omega]}t+\delta)}\\
p^\phi&=&\frac{A\Omega}{\sqrt{1-2M/r_0\;}} 
+\frac{A}{r_0}\cos{(\gamma^{-1}{[\Omega-\omega]}t+\delta)}\\
p^r&=&\gamma^{-1}{A}\sin{(\gamma^{-1}{[\Omega-\omega]}t+\delta)}\,.
\end{eqnarray}
This tells us that the photon starts out its journey with 
\begin{equation}
  \label{eq:drdphi}
  \frac{p^r}{p^\phi}=\frac{dr}{d\phi}
=\frac{r_0\gamma^{-1}
\sin({\gamma^{-1}[\Omega-\omega]} t+\delta)
}
{\frac{r_0\Omega}{\sqrt{1-2M/r_0\;}}
+ \cos{({\gamma^{-1}[\Omega-\omega]}t+\delta)}  }\,,
\end{equation}
with 
\begin{equation}
  \label{eq:drdt}
 \frac{p^r}{p^0}=\frac{dr}{dt}
=\frac{\gamma^{-1}
\sin({\gamma^{-1}[\Omega-\omega]} t+\delta)
}
{\frac{1}{\sqrt{1-2M/r_0\;}}
+\frac{r_0\Omega}{1-2M/r_0} \cos{({\gamma^{-1}[\Omega-\omega]}t+\delta)}  }\,,
 \end{equation}
and with 
\begin{equation}
  \label{eq:dphidt}
 \frac{p^\phi}{p^0}=\frac{d\phi}{dt}
=\frac
{\frac{\Omega}{\sqrt{1-2M/r_0\;}}
+\frac{1}{r_0} \cos{(\gamma^{-1}{[\Omega-\omega]}t+\delta)}  }
%%%%
{\frac{1}{\sqrt{1-2M/r_0\;}}
+\frac{r_0\Omega}{1-2M/r_0} \cos{(\gamma^{-1}{[\Omega-\omega]}t+\delta)}  }\,.
 \end{equation}
Equation (\ref{eq:tanphiin})
follows from Eq.~(\ref{eq:drdphi}) and from the definition of $\tan{\phi_{\rm in}}$
as the initial value of $r d\phi/dr$.

\section{Computation of the universal functions} 
\subsection{Integral for $\phi_\infty$}

For a photon beamed in the equatorial plane, we assume that we know its
initial radial location $r_0$ and its initial direction, as specified by the 
angle measured with respect to the outgoing radial direction
\begin{equation}
  \label{eq:phiindef}
 \phi_{\rm in}
=\tan^{-1}{\left(r_0d\phi/dr\right)} \,.
\end{equation}

The equation for the photon orbit is 
\begin{equation}
  \label{eq:photorbit2}
  \frac{1}{r^4}\left(\frac{dr}{d\phi}\right)^2+\frac{1-2M/r}{r^2}=\frac{1}{b^2}\ .
\end{equation}
From the known initial values of $r$ and of $dr/d\phi$ we solve
Eq.~(\ref{eq:photorbit2}) for the impact parameter $b$, a constant of
motion for the orbit. We proceed with the following steps to find
$\phi_\infty$. In the following we describe as ``delicate'' any
integral that has a divergent integrand, and for which special
techniques, desribed in Appendix~\ref{sec:integrals}, must be used.
\begin{enumerate}
\item Suppose $\cos{(\phi_{\rm in})}>0$, so that the photon is emitted
  going generally outward.  In this case the photon will be moving
  only to larger $r$, but it may be moving to larger or smaller
  $\phi$, and we must separate this case into two subcases depending
  on the initial value of $dr/d\phi$:
\begin{enumerate}
\item If initially $dr/d\phi>0$, then we use
\begin{equation}
  \label{eq:phiint}
  {\phi}_\infty(t)=\Omega t+\int_{r_0}^\infty\frac{dr}{
\sqrt{r^4/b(t)^2-r^2+2Mr
\;}}\ .
\end{equation}
\item If initially $dr/d\phi<0$ we use
\begin{equation}
  \label{eq:phiinttwo}
  {\phi}_\infty(t)=\Omega t-\int_{r_0}^\infty\frac{dr}{
\sqrt{r^4/b(t)^2-r^2+2Mr
\;}}\ .
\end{equation}
\end{enumerate}
In both these cases, the integrands do not diverge, and no special
techniques are needed to carry out the integration.
\item If $\cos{\phi_{\rm in}}=0$ it means that the photon is emitted on
  a trajectory tangent to the pulsar orbit. In this case we must check
  whether the the photon is going in the direction of increasing or
  decreasing $\phi$:
\begin{enumerate}
\item If   $\sin{\phi_{\rm in}}>0$, then 
we use Eq.~(\ref{eq:phiint}).
\item If  
$\sin{\phi_{\rm in}}<0$,
 then 
we use Eq.~(\ref{eq:phiinttwo}).
\end{enumerate}
In the 
$\cos{\phi_{\rm in}}=0$ 
case, either of the integrals is ``delicate'' since 
the integrand diverges at $r=r_0$. 
\item If 
$\cos{(\phi_{\rm in})}<0$
and if $b^2>27M^2$, then there will be a value of $r$,
less than $r_0$, at which $dr/d\phi=0$, i.e., at which the denominator of 
the integrand in Eq.~(\ref{eq:phiint}) vanishes. Call that value $r_{\rm min}$.
Then we need to consider the usual subcases:
\begin{enumerate}
\item If $\sin{(\phi_{\rm in})}>0$, then 
\begin{equation}
  \label{eq:inout}
  {\phi}_\infty(t)=\Omega t
+2\int_{r_{\rm min}}^{r_0}\frac{dr}{\sqrt{r^4/b(t)^2-r^2+2Mr\;}}
+\int_{r_0}^\infty\frac{dr}{\sqrt{r^4/b(t)^2-r^2+2Mr\;}}
\,.
\end{equation}
\item 
If  
$\sin{(\phi_{\rm in})}<0$,
 then 
\begin{equation}
  \label{eq:inouttwo}
  {\phi}_\infty(t)=\Omega t
-2\int_{r_{\rm min}}^{r_0}\frac{dr}{\sqrt{r^4/b(t)^2-r^2+2Mr\;}}
-\int_{r_0}^\infty\frac{dr}{\sqrt{r^4/b(t)^2-r^2+2Mr\;}}
\,.
\end{equation}
\end{enumerate}
\item If  $\cos{(\phi_{\rm in})}<0$
and if $b^2<27M$, then the
photon will be captured by the black hole and there is no meaning 
to ${\phi_\infty}$.
\end{enumerate}

\subsection{Integral for time to infinity}

We need to find the coordinate time it takes for a photon to reach
``infinity.''  The calculation must be divided into two subcases
depending on whether the photon starts going outward ($\cos{(\phi_{\rm
    in})}>0$) or inward ($\cos{(\phi_{\rm in})}<0$).  In both cases
the calculation is based on the equation for $dr/dt$ for the photon
motion (see MTW, Eqs.~25.64 and 25.66):
\begin{equation}
  \label{eq:dtdr}
  \frac{dt}{dr}=\frac{1}{b(1-2M/r)}\,
\frac{1}{\sqrt{\frac{1}{b^2}-\frac{1}{r^2}+\frac{2M}{r^3}\;}}\equiv J(r)\ .
\end{equation}

\begin{enumerate}
\item If the photon starts its trajectory at time $t_e$, 
with $\cos{(\phi_{\rm in})}>0$, then the photon will always 
be traveling outward and
  \begin{equation}
    \label{eq:TOAout}
    t_\infty=t_e+\int_{r_0}^\infty J(r) dr\,.
  \end{equation}
In this case the integral is not delicate. 
\item If the photon starts its trajectory with 
$\cos{(\phi_{\rm in})}=0$, then 
Eq.~(\ref{eq:TOAout})
again applies, but now the integral is delicate, since $f(r_0)$ is infinite.
\item If the photon starts its trajectory  with 
$\cos{(\phi_{\rm in})}<0$, then let $r_{\rm min}$ have 
the same meaning as in   Eq.~(\ref{eq:inout}) (\ref{eq:inouttwo}).
The time of arrival of the pulse is now given by
  \begin{equation}
    \label{eq:TOAin}
    t_\infty=t_e+2\int_{r_{\rm min}}^{r_0} J(r) dr+\int_{r_0}^\infty J(r) dr\,.
  \end{equation}
The first integral is delicate since $J(r_{\rm min})$ is infinite.
\end{enumerate}

\subsection{Computational details for integrals}\label{sec:integrals}
The  integrals for $\phi_{\infty}$ and $t_\infty$
cannot in general be expressed as elementary functions, and
must be handled by numerical methods. If there is no singularity in
the integrand, such as in the case $\cos(\phi_{\rm in})>0$, we use
adaptive Simpson quadrature with an absolute error limit set at
$10^{-6}$.

In the case that the integrand does diverge, the divergences occur at
the a root of the polynomial $ r^3-b^2r+2Mb^2 $. For $b^2<27M^2$ this
polynomial has a negative root and two postive roots. We denote the
largest root as $r_{\rm min}$ since it is the minimum radius the
photon trajectory will reach.  We then write
\begin{equation}
r^{4}/b^2-r^{2}+2Mr=(r-r_{\rm min})f(r)
\end{equation}
where $f={r^{2}}(r+r_{\rm min})/b^2-2M{r}/{r_{\rm min}}$
does not vanish in the intervals of integration that are considered.

The use of this is illustrated for one of the integrals occurring in 
Eqs.~(\ref{eq:inout}) and (\ref{eq:inouttwo}):
\begin{equation}
I\equiv\int_{r_{min}}^{r_{0}}\frac{dr}{\sqrt{r^{4}/b^{2}-r^{2}+2Mr}}
=\int_{r_{\rm min}}^{r_{0}}\frac{dr}{\sqrt{r-r_{\rm min}}\sqrt{f(r)}}
\end{equation}
$$
=\int_{r_{\rm min}}^{r_{0}}\frac{dr}{\sqrt{r-r_{\rm
      min}}\sqrt{f(r_{\rm min})}} +\int_{r_{\rm min}}^{r_{0}}\frac{dr}
{\sqrt{r-r_{\rm min}}}\left[\frac{1}{\sqrt{f(r)}}
  -\frac{1}{\sqrt{f(r_{\rm min})}}\right]
$$
$$
=\int_{r_{\rm min}}^{r_{0}}\frac{dr}{\sqrt{r-r_{\rm min}}
  \sqrt{f(r_{\rm min})}}+\int_{r_{\rm
    min}}^{r_{0}}\frac{dr}{\sqrt{r-r_{\rm min}}} \left[\frac{f(r_{\rm
      min})-f(r)}{\left(\sqrt{f(r_{\rm min})}
      +\sqrt{f(r)}\;\right)\sqrt{f(r_{\rm min})}\sqrt{f(r)}}\right]\ .
$$
The numerator 
in the second integrand is
\begin{equation}
f(r_{\rm min})-f(r)=-\frac{1}{b^{2}}(r-r_{\rm min})(r^{2}
+2r_{\rm min}r+2r_{\rm min}^{2}-2b^{2}M/r_{\rm min})\ .
\end{equation}
so that the evaluation of $I$ is  reduced to
\begin{equation}
  I=2\frac{\sqrt{r_{0}-r_{\rm min}}}{\sqrt{f(r_{\rm min})}}
  -\frac{1}{b^{2}\sqrt{f(r_{\rm min})}}\int_{r_{\rm min}}^{r_{0}}
  \frac{\sqrt{r-r_{\rm min}}(r^{2}+2r_{\rm min}r+2r_{\rm min}^{2}-2b^{2}M/r_{\rm min})dr}
  {(\sqrt{f(r_{\rm min})}+\sqrt{f(r)})\sqrt{f(r)}}\ .
\end{equation}
The remaining integral is nonsingular and can be evaluated e.g., with an
adaptive Simpson's routine.

An additional complication arises if $b^2$ is very close to $27M^2$. In 
this case the two positive roots of $f(r)$ approach each other, and 
the integral becomes very large. To deal with this case we introduce 
the notation $r_2$ for the smaller positive root and $r_3$
for the negative root, and we write
$$
I\equiv\int_{r_{\rm min}}^{r_{0}}\frac{dr}{\sqrt{r^{4}/b^{2}-r^{2}+2Mr}}
=b\int_{r_{\rm min}}^{r_{0}}\frac{dr}{\sqrt{r_{\rm min}(r_{\rm
      min}-r_3)(r-r_{\rm min})(r-r_2)}}\hspace{1in}
$$
$$
\hspace{1in}+b\int_{r_{\rm
    min}}^{r_{0}}\left[\frac{dr}{\sqrt{r(r-r_3)(r-r_{\rm
        min})(r-r_2)}}-\frac{dr}{\sqrt{r_{\rm min}(r_{\rm
        min}-r_3)(r-r_{\rm min})(r-r_2)}}\right]
$$
$$
=\frac{b}{\sqrt{r_{\rm min}(r_{\rm min}-r_3)}}
\left\{\int_{r_{\rm min}}^{r_{0}}\frac{dr}{\sqrt{(r-r_{\rm min})(r-r_2)}}
  -\int_{r_{\rm min}}^{r_{0}}\frac{\sqrt{r-r_{\rm min}}(r+r_{\rm min}-r_3)
    \;dr}{\left(\sqrt{r_{\rm min}(r_{\rm min}-r_3)}
    +\sqrt{r(r-r_3)}\,\right)\sqrt{r(r-r_3)(r-r_2)}}\right\}
$$
\begin{equation}\label{eq:dubdelicate}
  =\frac{b}{\sqrt{r_{\rm min}(r_{\rm min}-r_3})}
  \left\{2\log\left(\frac{\sqrt{r_{0}-r_{\rm min}}
        +\sqrt{r_{0}-r_2}}{\sqrt{r_{\rm min}-r_2}}\right)
    -\int_{r_{\rm min}}^{r_{0}}\frac{\sqrt{r-r_{\rm min}}(r+r_{\rm min}-r_3)
      \;dr}{\left(\sqrt{r_{\rm min}(r_{\rm min}-r_3)}+\sqrt{r(r-r_3)}\right)
      \sqrt{r(r-r_3)(r-r_2)}}\right\}
\end{equation}
The remaining integrand is well behaved at $r=r_{\rm min}$
and is straightforward to evaluate numerically.

To get asymptotic approximations for $\phi_{\rm in}$ near $\phi_{\rm
  crit}$ we take $b^2=27M^2+\epsilon$, and we find that $r_{\rm
  min}\approx3M+\sqrt{\epsilon\;}/3$ and
$r_{2}\approx3M-\sqrt{\epsilon\;}/3$.  If we set $\phi_{\rm
  in}=\phi_{\rm crit}-\Delta\phi$, and assume positive $\phi_{\rm in}$
then from Eqs.~(\ref{eq:phineq}) and (\ref{eq:crit}) we find that
$\Delta\phi\propto\epsilon$ and hence $\Delta\phi\propto (r_{\rm
  min}-r_2 )^2.$ The logarithmic dependence in
Eq.~(\ref{eq:dubdelicate}) becomes
\begin{equation}
  \log(1/\sqrt{ r_{\rm min}-r_2   \;})\approx \log{\Delta\phi}
\approx -\textstyle{\frac{1}{4}}\log(1-\phi_{\rm in}/\phi_{\rm crit})\ .
\end{equation}
The prefactor of the logarithm in Eq.~(\ref{eq:dubdelicate}) is
evaluated to 2 when the approximations $b=3\sqrt{b}$, $r_{\rm
  min}=r_2=3M$, $r_3=-6M$ are used, and we get the asymptotic approximation
$-\frac{1}{2}\log{(1- \phi_{\rm in}/\phi_{\rm crit} )}$. The first
term in Eq.~(\ref{Fapprox}) follows from using this asymptotic
approximation for Eq.~(\ref{eq:dubdelicate}) in Eq.~(\ref{eq:inout})
and
modifying the result to make it an odd
function of $\phi_{\rm in}$. The remaining terms are added to
Eq.~(\ref{Fapprox}) to give $F(\phi_{\rm in})\approx \phi_{\rm in}$
for small $\phi_{\rm in}$. Equation ~(\ref{Fapprox}), therefore, gives
the right qualitative behavior of $F(\phi_{\rm in})$ both for
$\phi_{\rm in}\rightarrow0$ and $\phi_{\rm in}\rightarrow \phi_{\rm
  crit} $.

The approximation for $T(\phi_{\rm in})$ in Eq.~(\ref{Fapprox})
follows from similar considerations with the addition of a Roemer time
delay $r_0(1-\cos{\phi_{\rm in}})$, the time it takes the pulsar
signal (with $c=1$) to cross the orbit on its way to the receiver.
The appearance of this effect is clear in the $r_0=100M$ curve 
for $T(\phi_{\rm in})$ in Fig.~\ref{fig:univfuns}.

It has been useful to have more accurate approximations to the
universal functions than the expressions in Eqs.~(\ref{Fapprox}) and
(\ref{Tapprox}). To that end we have fit the errors (differences from
the numerical integrals) in Eqs.~(\ref{Fapprox}) and (\ref{Tapprox})
with even-order Chebyshev polynomials. The following polynomials give
better than 1\% fits to the numerical calculations when used with the
coefficients in  Table~\ref{tab:table2}:
\begin{equation}
  F(\phi_{\rm in})=-{\rm sign}(\phi_{\rm in})
  \left[\log(1-\frac{|\phi_{\rm in}|}{\phi_{\rm crit}})+f_{2}\phi_{\rm in}^{2}
  +f_{4}\phi_{\rm in}^{4}+f_{6}\phi_{\rm in}^{6}+f_{8}\phi_{\rm in}^{8}\right]
  -\frac{\phi_{\rm in}}{\phi_{\rm crit}}+\phi_{\rm in}
\end{equation}

\begin{equation}
  T(\phi_{\rm in})=r_{0}(1-\cos{\phi_{\rm in}})-{3\sqrt{3}}\;
  \log(1-\frac{|\phi_{\rm in}|}{\phi_{\rm crit}})
  -\frac{3\sqrt{3}}{\phi_{\rm crit}}\;|\phi_{\rm in}|
  +t_{2}\phi_{\rm in}^{2}+t_{4}\phi_{\rm in}^{4}+t_{6}\phi_{\rm in}^{6}
  +t_{8}\phi_{\rm in}^{8}+t_{10}\phi_{\rm in}^{10}\,.
\end{equation}

\begin{deluxetable}{c@{\;}c@{\ \ }c@{\ \ }c@{\ \ }c@{\ \ }c@{\ \ }c@{\ \ }c@{\ \ }c@{\ \ }c@{\ \ }c}
\tabletypesize{\footnotesize}\tablewidth{0pt}
\tablecaption{\label{tab:table2}Coefficients of residual of universal
function F and T}
\startdata\hline\hline
$r_{0}(M)$ &$\phi_{\rm crit}$ &$f_{2}$ &$f_{4}$ &$f_{6}$ &$f_{8}$
&$t_{2}$ &$t_{4}$ &$t_{6}$ &$t_{8}$ &$t_{10}$ \\
\hline 5& 2.0895 & 0.0634022 &-0.0707841 &0.0160038 &-0.000411158
&-0.86385 &0.930642 &-0.160082 &0.00547166 &0.0\\
10& 2.6109 & 0.0171645 &0.00324181 &-0.000180664 &0.000315048
&-0.263976 &0.599057 &-0.115977 &0.00609665 &0.0\\
30& 2.9677 & -0.0352832 &0.0569467 &-0.0110258 &0.000910931
&-0.414603 &0.786804 &-0.212809 &0.0236697 &-0.00100843\\
100& 3.0896 & -0.128706 &0.134169 &-0.0274787 &0.00192051
&-0.884133 &0.929001 &-0.23247 &0.0249892 &-0.00103073\\
\hline
\enddata
\end{deluxetable}

\bibliographystyle{apj}
\bibliography{equatorial}

\end{document}